%% file: Lattice2017_44.tex
\documentclass[epj]{webofc}
\usepackage[utf8]{inputenc}
\usepackage[varg]{txfonts}   
\usepackage{booktabs}
\usepackage{xcolor}
\usepackage{multirow}
\definecolor{darkred}{rgb}{0.4,0.0,0.0}
\definecolor{darkgreen}{rgb}{0.0,0.4,0.0}
\definecolor{darkblue}{rgb}{0.0,0.0,0.4}
\usepackage[bookmarks,linktocpage,colorlinks,
    linkcolor = darkred,
    urlcolor  = darkblue,
    citecolor = darkgreen]{hyperref}
\usepackage{subfig}

\wocname{EPJ Web of Conferences}
\woctitle{Lattice2017}

\newcommand{\Dov}{\ensuremath{D_\text{ov}}}
\newcommand{\K}{\ensuremath{K_{x,\mu}}}

\newcommand{\B}{\ensuremath{B}}
\newcommand{\mq}{\ensuremath{m_q}}

\newcommand{\ja}{\ensuremath{j^5}}
\newcommand{\jxa}{\ensuremath{j_{x,\mu}^5}}
\newcommand{\scsc}{\ensuremath{\sigma_{\text{\tiny CSE}}}}
\newcommand{\scscf}{\ensuremath{\sigma^{\text{\tiny 0}}_{\text{\tiny CSE}}}}

\newcommand{\MeV}{\ensuremath{\text{MeV}}}

\newcommand{\MF}{\ensuremath{\Phi_B}}
\newcommand{\fermi}{\ensuremath{\text{fm}}}
\newcommand{\LS}{\ensuremath{L_S}}
\newcommand{\LT}{\ensuremath{L_T}}
\newcommand{\gpgg}{\ensuremath{ g_{\pi^0\gamma\gamma}}}
\newcommand{\bzero}{\ensuremath{\beta}}

\newcommand{\Tc}{\ensuremath{\text{T}_\text{c}}} 
\DeclareMathOperator{\Tr}{tr}
\DeclareMathOperator{\Su}{SU}

\newcommand{\lr}[1]{ \left( #1 \right) }

\begin{document}
%
\selectlanguage{english}
\title{%
The Chiral Separation Effect in quenched finite-density QCD
}
\author{
 \firstname{Matthias} \lastname{Puhr}\inst{1}\fnsep\thanks{Speaker,
 \email{matthias.puhr@physik.uni-regensburg.de}} 
 \and 
 \firstname{Pavel}
 \lastname{Buividovich}\inst{1}
}
\institute{%
Institute of Theoretical Physics, Regensburg University, 93040 Regensburg, Germany 
}
\abstract{%
  We present results of a study of the Chiral Separation Effect (CSE) in quenched
 finite-density QCD. Using a recently developed numerical method we calculate the conserved
 axial current for exactly chiral overlap fermions at finite density for the first time. We
 compute the anomalous transport coefficient for the CSE in the confining and de-confining
 phase and investigate possible deviations from the universal value. In both phases we find
 that non-perturbative corrections to the CSE are absent and we reproduce the universal
 value for the transport coefficient within small statistical errors. Our results suggest
 that the CSE can be used to determine the renormalisation factor of the axial current.
}
\maketitle
\section{Motivation and Introduction}\label{sec:intro}

In heavy-ion collision experiments it is possible to generate densities and temperatures
that are comparable to the conditions in the early universe. These experiments are an
important tool to study open questions in cosmology, astrophysics and high-energy
physics. The hot and dense plasma generated in heavy-ion collisions is dominated by quarks
and gluons. QCD is an asymptotically free theory and at high enough temperatures and
densities one expects that the quarks and gluons become deconfined. If the collisions are
off-centre very large magnetic fields can be
generated~\cite{Kharzeev:08:01,Skokov:09:01,McLerran:13:01}. For these reasons one can
expect that anomalous transport phenomena~\cite{Kharzeev:16:01, Liao:16:01} might play a
role in heavy-ion collision experiments and it is of great interest to study anomalous
transport in QCD.

Prominent examples of anomalous transport phenomena are the induction of an axial or
vector current parallel to an external magnetic field in a dense chiral medium, the
so-called Chiral Separation effect (CSE)~\cite{Son:04:01,
Metlitski:05:01,Son:07:01,Kharzeev:07:01} and Chiral Magnetic effect
(CME)~\cite{Vilenkin:80:01,Fukushima:08:01}, respectively. In combination the CSE and the
CME can give rise to a gap-less hydrodynamic mode, the Chiral Magnetic
Wave~\cite{Burnier:11:01,Kharzeev:11:02}. For reviews about the experimental
signatures of anomalous transport effects see for example~\cite{Kharzeev:16:01,
Liao:16:01}.

Because of their relation to the axial anomaly it has been argued that the anomalous
transport coefficients are universal and do not get any corrections in interacting
theories. Closer investigations revealed, however, that there are two scenarios where
corrections to the anomalous transport coefficients can occur: If chiral symmetry is
spontaneously broken~\cite{Newman:06:01, Buividovich:13:02, Buividovich:14:02} and in an
unquenched theory if the currents couple to dynamical gauge
fields~\cite{Jensen:13:01,Gorbar:13:01,Gursoy:14:01}.

The focus of this work is the CSE in QCD, where non-perturbative corrections to the
transport coefficient can be expressed in terms of the in-medium amplitude $\gpgg$ of the
decay $\pi^0 \to \gamma \gamma$~\cite{Newman:06:01}:
\begin{equation}
\label{eq:chiralsep} \ja_i = \scsc B_i, \quad \scsc= \scscf \lr{1 - g_{\pi^0 \gamma\gamma}} ,
\end{equation}
where  $\ja_i$ is the axial current density  and $\B_i$ the external magnetic
field. In the limit $\gpgg \to 0$ the transport coefficient $\scsc$  reduces to the value
for free chiral quarks $\scscf$. For a single quark flavour with $N_c$ colour degrees of
freedom it is given by
\begin{equation}
  \label{eq:scscf} \scscf = \frac{q N_c \mu}{2 \pi^2},
\end{equation} where $q$ is the electrical charge of the quark and $\mu$ the quark
chemical potential.

In the linear sigma model $\gpgg$ can be calculated and in the phase with broken chiral
symmetry (for sufficiently small chemical potential) it is given by $\gpgg = \frac{7
\zeta\lr{3} m^2}{4 \pi^2 T^2}$, where $\zeta$ is the Riemann $\zeta$-function, $m$ is the
constituent quark mass and $T$ is the temperature~\cite{Newman:06:01}. Plugging in the
values $m \sim 300 \ \MeV$ and $T \sim 150 \ \MeV$, which give a realistic low-energy
description of the chirally broken phase of QCD~\cite{Baboukhadia:97:01}, we find a
correction of order $100 \%$ which suppresses the CSE current. Corrections suppressing the
CSE were also found in other model calculations~\cite{Gorbar:09:01, Gorbar:11:01,
Gorbar:11:02, Amado:14:01,Jimenez-Alba:14:01}.

For accurate predictions of signatures of anomalous transport effects in
heavy-ion collision experiments it is desirable to gain a quantitative, model-independent
understanding of possible corrections to the anomalous transport coefficients from
first-principle lattice QCD simulations. Previous lattice studies looked at the infrared
values of the anomalous transport coefficients for the
CME~\cite{Yamamoto:11:01,Yamamoto:11:02} and the Chiral Vortical Effect
(CVE)~\cite{Braguta:13:01,Braguta:14:01}. These studies found a significant suppression of
the CME and the CVE at both low and high temperatures, conflicting with expectations based
on the hydrodynamic approximation. At least at high temperatures the thermodynamic
consistency arguments fixing the anomalous transport coefficients within this
approximation should be valid~\cite{Son:09:01,Sadofyev:11:01}. It is possible that the
origin of this discrepancy lies in the use of a naively discretised non-conserved vector
current~\cite{Yamamoto:11:01,Yamamoto:11:02} and energy-momentum
tensor~\cite{Braguta:13:01,Braguta:14:01}. Moreover, the simulations
in~\cite{Yamamoto:11:01,Yamamoto:11:02} were performed with non-chiral Wilson--Dirac
lattice fermions.

In this contribution we report on a first-principles lattice study of the CSE, previously
published in~\cite{Puhr:16:03}. To avoid unquantifiable systematic errors we work with
finite-density overlap fermions~\cite{Bloch:06:01}, which respect a lattice version of
chiral symmetry, and use the properly defined conserved lattice axial vector current
density~\cite{Hasenfratz:02:01,Kikukawa:98:01}:
\begin{equation}
\label{eq:j5} \jxa = \tfrac{1}{2} \bar{\psi} \left( - \gamma_5\K + \K
\gamma_5(1-\Dov)\right) \psi ,
\end{equation} 
where $\K = \frac{\partial \Dov}{\partial \Theta_{x,\mu}}$ is the
derivative of the overlap operator $\Dov$ over the $U(1)$ lattice gauge field
$\Theta_{x,\mu}$. With the definition \eqref{eq:j5} the lattice axial current transforms
covariantly under the lattice chiral symmetry. For vanishing bare quark mass it is
therefore protected from renormalisation and can be directly related to the continuum
axial current density $j^5_\mu=\bar{\psi}\gamma_5\gamma_\mu\psi$, which enters
Equation~\eqref{eq:chiralsep}. Taking the expectation value of~(\ref{eq:j5}) and using the
Ginsparg--Wilson equation to simplify the resulting expression finally yields
\begin{equation}
  \label{eq:<j5>} \langle \jxa \rangle = \Tr\left(\Dov^{-1}\frac{\partial \Dov}{\partial
\Theta_{x,\mu}} \gamma_5 \right).
\end{equation} 
Efficiently computing the derivatives $\frac{\partial \Dov}{\partial
\Theta_{x,\mu}}$ with high accuracy is a non-trivial numerical problem and we developed a
new numerical algorithm for this purpose. For details on the evaluation of the derivatives
we refer the reader to~\cite{Puhr:16:01}.

\section{Simulation parameters and numerical setup}\label{sec:setup}

Lattice QCD with dynamical fermions has a sign problem at finite quark chemical
potential. In order to avert the sign problem we work in the quenched approximation and
neglect the effects of sea quarks. While calculations within a random matrix model show
that the chiral condensate in quenched QCD vanishes and chiral symmetry is restored for
any non-zero chemical potential~\cite{Stephanov:96:01}, the presence of an external
magnetic field can potentially change this picture. On the one hand random matrix theory
is no longer applicable in this case and on the other hand non-perturbative corrections to
the CSE due to the formation of a new type of condensate, the so-called chiral shift
parameter~\cite{Gorbar:09:01,Gorbar:11:01,Gorbar:11:02}, are possible.

The $\Su(3)$ gauge configurations are generated using the tadpole-improved Lüscher--Weisz
gauge action~\cite{Luescher:85:01}. We use three different parameter sets for our
simulations: $V=\LT\times \LS^3 = 6\times 18^3$ with $\bzero=8.45$ corresponding to a
temperature $T>T_c$ and $V = 14\times 14^3$ and $V = 8 \times 8^3$ with $\bzero=8.10$
corresponding to $T<T_c$, where $\LT$ and $\LS$ are the temporal and spatial extent of the
lattice and $T_c \approx 300 \ \MeV$ is the deconfinement transition temperature of the
Lüscher--Weisz action~\cite{Gattringer:02:01}. To fix the lattice spacing $a$ we take the
results from~\cite{Gattringer:01:01}. The values of all parameters in lattice and physical
units are summarised in Table~\ref{tab:params}.

\begin{table}[thb] 
\small
 \centering
    \begin{tabular}{p{3pt}llccc}
    \toprule[1pt]
     \multicolumn{2}{l}{\multirow{2}{*}{Setup}}   & $\bzero$ & $8.1$ & $8.1$ & $8.45$ \\
      & &  Volume & $14 \times 14^3 $ & $8 \times 8^3 $& $6 \times 18^3 $ \\
     \midrule[1pt]
    & & Lattice & \multicolumn{3}{c}{Physical Value} \\
    \midrule[1pt]
    $a$ & $[\fermi]$ & $1$ & $0.125$ & $0.125$ & $0.095$ \\
    $V_S $ & $ [\fermi^3]$ & $\LS^3$ & $5.4$ & $1.0$ & $5.0$ \\
    $T $ & $ [\MeV]$ & $\LT^{-1}$ & $113$ & $197$ & $346$ \\
    $\mu $ & $ [\MeV]$ & $0.050  $ &$79$ & $\cdots$ & $\cdots$ \\
    & & $0.100  $ & $\cdots$ & $158$ & $\cdots$ \\
    & & $0.300  $ & $474$ & $\cdots$ & $\cdots$ \\
    & & $0.040  $ & $\cdots$ & $\cdots$ & $83$ \\
    & & $0.230  $ & $\cdots$ & $\cdots$ & $478$ \\
    $\frac{qB}{\MF}$ & $[\MeV]^2$ & $\frac{2\pi}{a^2\LS^2}$ & $283^2 $ & $495^2$& $289^2 $ \\
    \bottomrule[1pt]
  \end{tabular}
  \caption{Simulation parameters}
  \label{tab:params}
\end{table}

For the $6 \times 18^3$ and $14 \times 14^3$ lattices approximately $10^3$ configurations
were generated, from which we randomly picked $100$ with topological charge
$Q=0$\footnote{One of the configurations for the parameters $V=14 \times 14^3$,
$\bzero=8.1$, $\mu = 0.050$ and a magnetic flux of $\MF = 1$ caused a serious breakdown in
the Lanczos algorithm when computing the overlap operator and only the remaining $99$
configurations were used for this parameter set.}. Additionally we chose $100$
configurations with topological charge $|Q| = 1$ for $V=6 \times 18^3$ and $111$
with $|Q|= 1$ and $97$ with $|Q|= 2$ for $V=14 \times 14^3$. For the $V=8 \times 8^3$ lattice
$5 \cdot 10^3$ configurations were generated, from which we selected three random sets of
$200$ configurations with $Q = 0$, $|Q| = 1$ and $|Q| = 2$.

The topological charge of a given gauge configuration can be calculated by taking the
difference of the number of left- and right-handed zero modes of the overlap operator:
\mbox{$Q= n_L- n_R$}. In practice configurations with zero modes with both chiralities do
not occur and the overlap operator always has either $n_R = |Q|$ right-handed or $n_L =
|Q|$ left-handed zero modes (see e.g. Section~7.3.2 in
\cite{GattringerLATTICE_QCD}). Exploiting this fact we calculated the absolute value of
the topological charge $|Q| = |n_R - n_L|$ as the number of zero eigenvalues of the
operator $\Dov \Dov^{\dag}$.

To introduce a constant, homogeneous external magnetic field on the lattice we
follow~\cite{Al_Hashimi:09:01} and introduce a magnetic flux quantum $\MF = 1, 2, 5,10$
for $V = 14 \times 14^3$ and $V = 6 \times 18^3$ at $Q=0$, and $\MF = 0, 1,2,3,4$ for $V =
8 \times 8^3$ at all $Q$.  For $V = 6 \times 18^3$ we chose $\MF = 0, 1,2,3,5$ at $|Q|= 1$
and $\MF = 1,3,5,8,10$ for $V = 14 \times 14^3$ at $|Q|= 1, 2$. The axial current density
is computed by averaging (\ref{eq:<j5>}) over all lattice sites $x$. To evaluate the trace
we use the stochastic estimator technique with $Z_2$-noise. The number of stochastic
estimators is increased until the results are stable. The axial current
density is only well defined if the overlap operator is invertible, i.e., if
$Q=0$. Working exclusively on configurations with $Q=0$ introduces an systematic error and
in order to perform a cross-check of our results we also consider configurations with
$|Q|>0$. Since the computations are numerically very expensive, we only do the
cross-checks for a single value of the chemical potential. By introducing a small finite
quark mass $m_q = 0.001 \ a^{-1}$ on configurations with non-zero topological charge we
make the overlap operator invertible. Strictly speaking the axial current defined via
Equation~(\ref{eq:<j5>}) is no longer protected from renormalisation in this case. To
demonstrate that the effect of the finite quark mass on $\scsc$ is negligible in practice,
we consider a second mass value $m_q = 0.002 \ a^{-1}$ for the $V = 8 \times 8^3$
configurations.

The value of $\scsc$ is given by the slope of the axial current density as a function of
the external magnetic field. We extract $\scsc$ from our axial current data by performing
a one parameter linear fit. Confidence intervals for $\scsc$ are calculated with the
statistical bootstrap method: For every bootstrap sample we first independently draw $100$
configurations for every value of $\MF$ and then perform a fit to the data generated in
this way.

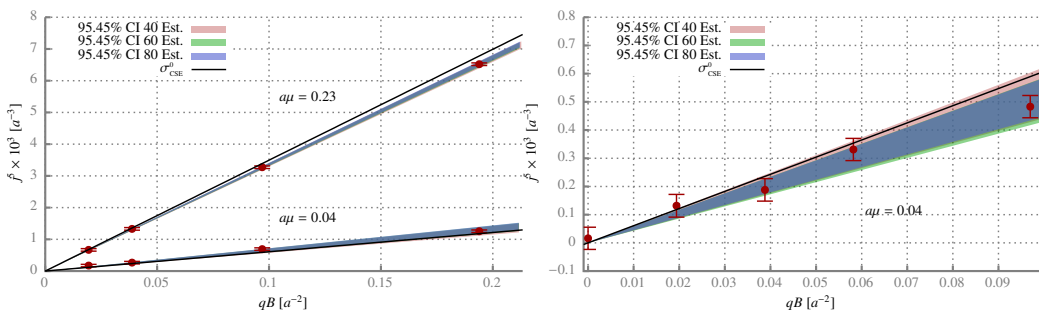
\begin{figure}[htb]
  \centering
  \resizebox{0.48 \textwidth}{!}{\input{6x18b845_slope_CI}}%
  \resizebox{0.48 \textwidth}{!}{\input{6x18b845_slope_CI_Q01}}%
  \caption{The axial current density $\ja$ as a function of the magnetic field strength
            $B$ for $T < \Tc$. The left plot shows results for $Q=0$ and on the right
            $|Q|=1$ (note the different scales). The red dots with errorbars are our data
            and the shaded regions mark the bootstrap confidence intervals for $\scsc$ for
            a different number of stochastic estimators. Solid black lines correspond to
            the free fermion result $\scscf$.  }
 \label{fig:cse_6x18_CI}
\end{figure}

\section{Results}\label{sec:res}
First we present results for the high-temperature deconfinement phase, where
$T~=~346~\MeV~>~T_c$. Here the chiral symmetry should be restored\footnote{The restoration
of chiral symmetry in the deconfinement phase of quenched lattice QCD is discussed, e.g.,
in~\cite{Edwards:99:01,Kiskis:01:01}} and we expect that there are no corrections to the
CSE current~\cite{Alekseev:98:01,Metlitski:05:01,Newman:06:01}. Our data is plotted in
Figure~\ref{fig:cse_6x18_CI} and as expected in general we find good agreement with the
free fermion result $\scscf$. The sole exception is the data point for $Q=0$, $\mu = 0.230
\ a^{-1}$ and $\MF=10$, where we might see the onset of saturation. To make sure that our
results for $\scsc$ are not affected by possible statuartion effects, we additionally
perform fits where the data for the largest value of $\MF$ is left out (see
Figure~\ref{subfig:CIs}).

\begin{figure}[!ht]
  \subfloat[
    Summary of the results for the confidence intervals for the ratio $\scsc/\scscf$ for
    the lattices with $V=14\times14^3$ and $V=6\times18^3$. Results for $T>\Tc$ and
    $T<\Tc$ are marked by open and closed boxes, respectively. The boxes denote the
    results of a fit to all data points and the whiskers show the results if the data for
    the largest value of $\MF$ are excluded.\label{subfig:CIs}]
    {%
      \resizebox{0.48\textwidth}{!}{\input{CI_graph}}%
    }
    \hfill
    \subfloat[
    The axial current density $\ja$ in different topological sectors for the
    $V=8\times8^3$ lattice. The results for $m_q = 0.001 \ a^{-1}$ are denoted by filled
    symbols, the data for $m_q = 0.002 \ a^{-1}$ are shifted by $0.02 \ a^{-2}$ in the
    $qB$ axis for better visibility and are marked by open symbols. The black dots show
    the axial current with $Q = 0$ for vanishing quark mass and the black dashed line
    corresponds to the free fermion result $\scscf$. To guide the eye a linear ($Q=0$) or
    second order polynomial ($|Q|>0$) fit to the data is shown.\label{subfig:8x8}]
    {%
      \resizebox{0.48\textwidth}{!}{\input{8x8b81_mu0100_all_mo}}%
    }
    \caption{}
\end{figure}

Next we examine the results for the low-temperature confinement phase, where
non-perturbative corrections to the CSE are expected. As a proof of concept we first
consider the small $V=8\times8$ lattice. In the topological sector $Q=0$ we
again find a very good agreement with $\scscf$, but for $|Q| \neq 0$ there are large
deviations from the free fermion result. The results for different bare quark masses 
lie on top of each other and we conclude that for small quark masses the
renormalisation of the axial current is negligible.

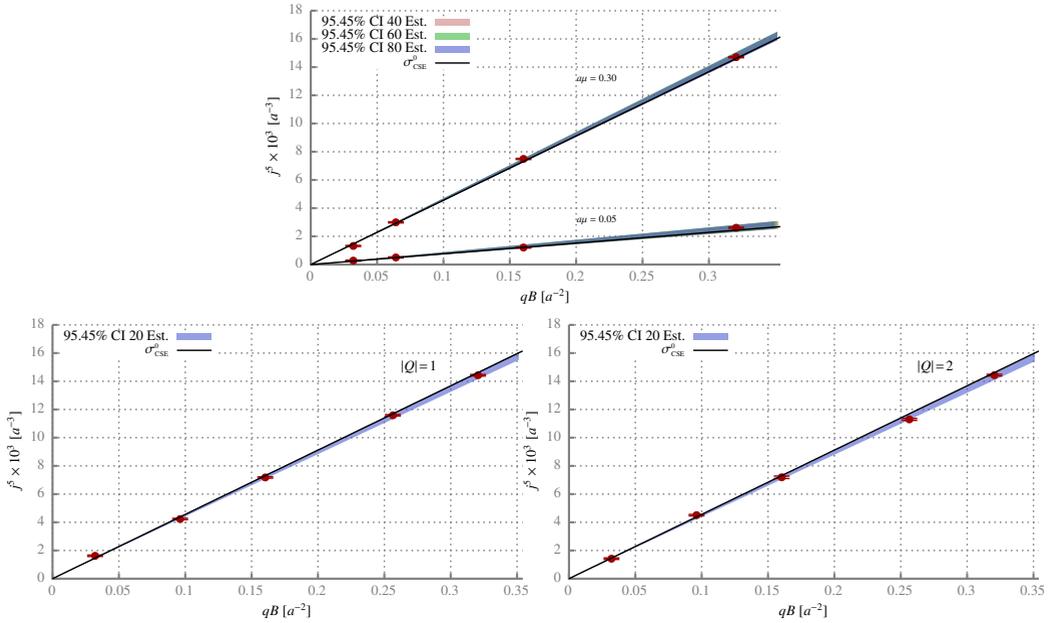
\begin{figure}[htb]
  \centering
  \resizebox{0.48 \textwidth}{!}{\input{14x14b81_slope_CI}} \\ %
  \resizebox{0.48 \textwidth}{!}{\input{14x14b81_mu0300_slope_CI_Q01}}%
  \resizebox{0.48 \textwidth}{!}{\input{14x14b81_mu0300_slope_CI_Q02}}%
  \caption{The axial current density $\ja$ as a function of the magnetic field strength
           $B$ in different topological sectors for the lattice with $V=14\times14^3$ at
           $T <  \Tc$ (red dots with errorbars).  In the plot on the top $Q=0$,
           $|Q|=1$ on the bottom left plot and on the bottom right $|Q|=2$. Solid black
           lines denote the free fermion result $\scscf$ and shaded regions mark the
           confidence intervals for $\scsc$.}
  \label{fig:cse_14x14_CI_Q1_Q2}
\end{figure}

A lattice volume of $V=8\times8^3$ is very small and to check for finite size effects we
also perform simulations for $V=14\times14^3$. The results for the larger lattice are
shown in Figure~\ref{fig:cse_14x14_CI_Q1_Q2}. The data for $Q=0$ is in very good agreement
with the results for the smaller lattice size and there does not seem to be a large finite
size effect for this topological sector. For the $|Q| \neq 0$ sectors the picture is
completely different: Contrary to the small volume calculations the CSE current does not
get any corrections. The plots in Figure~\ref{fig:cse_14x14_CI_Q1_Q2} clearly show that
for the larger lattice volume the data for all topological sectors and chemical potentials
we investigated are in perfect agreement with the free fermion result $\scscf$. A possible
reason for the large finite size effects in topological sectors with non-zero $Q$ is
discussed in~\cite{Puhr:16:03}. The results for all our simulations on larger lattices are
summarised in Figure~\ref{subfig:CIs}.

\section{Conclusion}\label{sec:conclusion}
We performed a numerical study to quantify possible non-perturbative corrections to the
CSE current in quenched lattice QCD. Within statistical errors, which are smaller than
$10\%$ for the simulations with larger chemical potentials (see Figure~\ref{subfig:CIs}),
we do not find any correction to the CSE current and reproduce the free fermion value
$\scscf$ for the transport coefficient. The use of finite-density overlap fermions and a
conserved lattice axial current, which transforms covariantly under the lattice chiral
symmetry, eliminates potential systematic errors due to a explicit breaking of chiral
symmetry or a renormalisation of the axial current. Comparing the results for different
lattice sizes suggest that finite size effects are very small, at least in the
topological sector $Q=0$. A remaining source of systematic errors is the quenched
approximation. Taking the results of the random matrix model~\cite{Stephanov:96:01} at
face value, one could argue that the chiral condensate in quenched QCD should vanish as
soon as a finite chemical potential is turned on and consequently the non-perturbative
corrections predicted for the phase with broken chiral symmetry should be absent. However,
on the one hand the random matrix calculation can not take into account a finite external
magnetic field and on the other hand the presence of such a field can instigate the
spontaneous formation of condensates, like the chiral shift parameter
of~\cite{Gorbar:09:01,Gorbar:11:01,Gorbar:11:02}, which can also give non-perturbative
corrections to the CSE. Moreover, the holographic
calculations~\cite{Amado:14:01,Jimenez-Alba:14:01} found non-perturbatice corrections to
the CSE at small temperatures and were done in the quenched approximation (or ``probe
limit'' in the language of AdS/CFT). For all this reasons the non-renormalisation of the
CSE current in quenched QCD at both high and low temperatures is a non-trivial result.

It is important to emphasise that the results for the quenched theory do not necessarily
generalise to full QCD. In particular, our results do not exclude possible corrections that
could have their origin in the complex phase that the fermion determinate acquires at
finite chemical potential. Note that unquenched lattice calculations of the CSE are
notoriously difficult, since the necessity to introduce an external magnetic field leads
to a complex fermion determinant even in gauge theories which do otherwise not have a sign
problem at finite chemical potential, like for example $\Su(2)$ and $G_2$ gauge theories. 

The non-renormalisation of the CSE in quenched QCD has a potential practical
application: If the axial current is computed for non-chiral lattice fermions and/or with
a non-covariant discretisation of the axial current, the ratio of this current and the
exact result $\ja_i = \scscf B_i$ gives the multiplicative renormalisation constant for
the axial current for this particular lattice discretisation of the Dirac operator and
axial current. 

\section*{Acknowledgements}\label{sec:ack}
This work was supported by the S.~Kowalevskaja award from the
Alexander von Humboldt Foundation. The computations were performed on
``iDataCool'' at Regensburg University, on the ITEP cluster in Moscow and on
the LRZ cluster in Garching. We thank  G.~Bali, A.~Dromard,  R. Rödl and A.~Zhitnitsky for
valuable discussions and helpful comments.

\end{document}

%% file: 6x18b845_slope_CI
\begingroup
  \makeatletter
  \providecommand\color[2][]{%
    \GenericError{(gnuplot) \space\space\space\@spaces}{%
      Package color not loaded in conjunction with
      terminal option `colourtext'%
    }{See the gnuplot documentation for explanation.%
    }{Either use 'blacktext' in gnuplot or load the package
      color.sty in LaTeX.}%
    \renewcommand\color[2][]{}%
  }%
  \providecommand\includegraphics[2][]{%
    \GenericError{(gnuplot) \space\space\space\@spaces}{%
      Package graphicx or graphics not loaded%
    }{See the gnuplot documentation for explanation.%
    }{The gnuplot epslatex terminal needs graphicx.sty or graphics.sty.}%
    \renewcommand\includegraphics[2][]{}%
  }%
  \providecommand\rotatebox[2]{#2}%
  \@ifundefined{ifGPcolor}{%
    \newif\ifGPcolor
    \GPcolortrue
  }{}%
  \@ifundefined{ifGPblacktext}{%
    \newif\ifGPblacktext
    \GPblacktextfalse
  }{}%
  \let\gplgaddtomacro\g@addto@macro
  \gdef\gplbacktext{}%
  \gdef\gplfronttext{}%
  \makeatother
  \ifGPblacktext
    \def\colorrgb#1{}%
    \def\colorgray#1{}%
  \else
    \ifGPcolor
      \def\colorrgb#1{\color[rgb]{#1}}%
      \def\colorgray#1{\color[gray]{#1}}%
      \expandafter\def\csname LTw\endcsname{\color{white}}%
      \expandafter\def\csname LTb\endcsname{\color{black}}%
      \expandafter\def\csname LTa\endcsname{\color{black}}%
      \expandafter\def\csname LT0\endcsname{\color[rgb]{1,0,0}}%
      \expandafter\def\csname LT1\endcsname{\color[rgb]{0,1,0}}%
      \expandafter\def\csname LT2\endcsname{\color[rgb]{0,0,1}}%
      \expandafter\def\csname LT3\endcsname{\color[rgb]{1,0,1}}%
      \expandafter\def\csname LT4\endcsname{\color[rgb]{0,1,1}}%
      \expandafter\def\csname LT5\endcsname{\color[rgb]{1,1,0}}%
      \expandafter\def\csname LT6\endcsname{\color[rgb]{0,0,0}}%
      \expandafter\def\csname LT7\endcsname{\color[rgb]{1,0.3,0}}%
      \expandafter\def\csname LT8\endcsname{\color[rgb]{0.5,0.5,0.5}}%
    \else
      \def\colorrgb#1{\color{black}}%
      \def\colorgray#1{\color[gray]{#1}}%
      \expandafter\def\csname LTw\endcsname{\color{white}}%
      \expandafter\def\csname LTb\endcsname{\color{black}}%
      \expandafter\def\csname LTa\endcsname{\color{black}}%
      \expandafter\def\csname LT0\endcsname{\color{black}}%
      \expandafter\def\csname LT1\endcsname{\color{black}}%
      \expandafter\def\csname LT2\endcsname{\color{black}}%
      \expandafter\def\csname LT3\endcsname{\color{black}}%
      \expandafter\def\csname LT4\endcsname{\color{black}}%
      \expandafter\def\csname LT5\endcsname{\color{black}}%
      \expandafter\def\csname LT6\endcsname{\color{black}}%
      \expandafter\def\csname LT7\endcsname{\color{black}}%
      \expandafter\def\csname LT8\endcsname{\color{black}}%
    \fi
  \fi
    \setlength{\unitlength}{0.0500bp}%
    \ifx\gptboxheight\undefined%
      \newlength{\gptboxheight}%
      \newlength{\gptboxwidth}%
      \newsavebox{\gptboxtext}%
    \fi%
    \setlength{\fboxrule}{0.5pt}%
    \setlength{\fboxsep}{1pt}%
\begin{picture}(7200.00,4320.00)%
    \gplgaddtomacro\gplbacktext{%
      \colorrgb{0.50,0.50,0.50}%
      \put(441,595){\makebox(0,0)[r]{\strut{}$0$}}%
      \colorrgb{0.50,0.50,0.50}%
      \put(441,1035){\makebox(0,0)[r]{\strut{}$1$}}%
      \colorrgb{0.50,0.50,0.50}%
      \put(441,1475){\makebox(0,0)[r]{\strut{}$2$}}%
      \colorrgb{0.50,0.50,0.50}%
      \put(441,1915){\makebox(0,0)[r]{\strut{}$3$}}%
      \colorrgb{0.50,0.50,0.50}%
      \put(441,2355){\makebox(0,0)[r]{\strut{}$4$}}%
      \colorrgb{0.50,0.50,0.50}%
      \put(441,2795){\makebox(0,0)[r]{\strut{}$5$}}%
      \colorrgb{0.50,0.50,0.50}%
      \put(441,3235){\makebox(0,0)[r]{\strut{}$6$}}%
      \colorrgb{0.50,0.50,0.50}%
      \put(441,3675){\makebox(0,0)[r]{\strut{}$7$}}%
      \colorrgb{0.50,0.50,0.50}%
      \put(441,4115){\makebox(0,0)[r]{\strut{}$8$}}%
      \colorrgb{0.50,0.50,0.50}%
      \put(543,409){\makebox(0,0){\strut{}$0$}}%
      \colorrgb{0.50,0.50,0.50}%
      \put(2103,409){\makebox(0,0){\strut{}$0.05$}}%
      \colorrgb{0.50,0.50,0.50}%
      \put(3663,409){\makebox(0,0){\strut{}$0.1$}}%
      \colorrgb{0.50,0.50,0.50}%
      \put(5223,409){\makebox(0,0){\strut{}$0.15$}}%
      \colorrgb{0.50,0.50,0.50}%
      \put(6783,409){\makebox(0,0){\strut{}$0.2$}}%
      \csname LTb\endcsname%
      \put(3819,1317){\makebox(0,0)[l]{\strut{}$a\mu = 0.04$}}%
      \csname LTb\endcsname%
      \put(3819,2989){\makebox(0,0)[l]{\strut{}$a\mu = 0.23$}}%
    }%
    \gplgaddtomacro\gplfronttext{%
      \csname LTb\endcsname%
      \put(144,2355){\rotatebox{-270}{\makebox(0,0){\strut{}$\ja  \times 10^{ 3}\;[a^{-3}]$}}}%
      \csname LTb\endcsname%
      \put(3871,130){\makebox(0,0){\strut{}$qB\;[a^{-2}]$}}%
      \csname LTb\endcsname%
      \put(2481,3948){\makebox(0,0)[r]{\strut{}$95.45\%$ CI $40$ Est.}}%
      \csname LTb\endcsname%
      \put(2481,3762){\makebox(0,0)[r]{\strut{}$95.45\%$ CI $60$ Est.}}%
      \csname LTb\endcsname%
      \put(2481,3576){\makebox(0,0)[r]{\strut{}$95.45\%$ CI $80$ Est.}}%
      \csname LTb\endcsname%
      \put(2481,3390){\makebox(0,0)[r]{\strut{}$\scscf$}}%
    }%
    \gplbacktext
    \put(0,0){\includegraphics{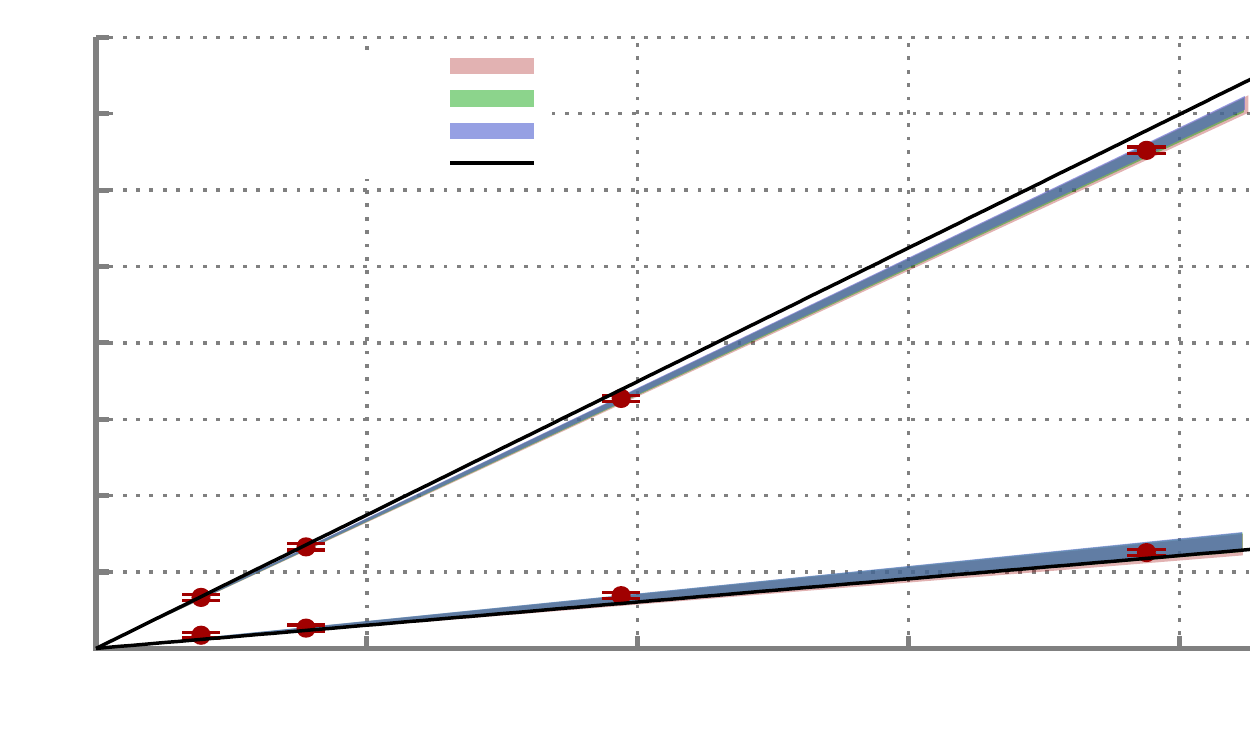}}%
    \gplfronttext
  \end{picture}%
\endgroup

%% file: 6x18b845_slope_CI_Q01
\begingroup
  \makeatletter
  \providecommand\color[2][]{%
    \GenericError{(gnuplot) \space\space\space\@spaces}{%
      Package color not loaded in conjunction with
      terminal option `colourtext'%
    }{See the gnuplot documentation for explanation.%
    }{Either use 'blacktext' in gnuplot or load the package
      color.sty in LaTeX.}%
    \renewcommand\color[2][]{}%
  }%
  \providecommand\includegraphics[2][]{%
    \GenericError{(gnuplot) \space\space\space\@spaces}{%
      Package graphicx or graphics not loaded%
    }{See the gnuplot documentation for explanation.%
    }{The gnuplot epslatex terminal needs graphicx.sty or graphics.sty.}%
    \renewcommand\includegraphics[2][]{}%
  }%
  \providecommand\rotatebox[2]{#2}%
  \@ifundefined{ifGPcolor}{%
    \newif\ifGPcolor
    \GPcolortrue
  }{}%
  \@ifundefined{ifGPblacktext}{%
    \newif\ifGPblacktext
    \GPblacktextfalse
  }{}%
  \let\gplgaddtomacro\g@addto@macro
  \gdef\gplbacktext{}%
  \gdef\gplfronttext{}%
  \makeatother
  \ifGPblacktext
    \def\colorrgb#1{}%
    \def\colorgray#1{}%
  \else
    \ifGPcolor
      \def\colorrgb#1{\color[rgb]{#1}}%
      \def\colorgray#1{\color[gray]{#1}}%
      \expandafter\def\csname LTw\endcsname{\color{white}}%
      \expandafter\def\csname LTb\endcsname{\color{black}}%
      \expandafter\def\csname LTa\endcsname{\color{black}}%
      \expandafter\def\csname LT0\endcsname{\color[rgb]{1,0,0}}%
      \expandafter\def\csname LT1\endcsname{\color[rgb]{0,1,0}}%
      \expandafter\def\csname LT2\endcsname{\color[rgb]{0,0,1}}%
      \expandafter\def\csname LT3\endcsname{\color[rgb]{1,0,1}}%
      \expandafter\def\csname LT4\endcsname{\color[rgb]{0,1,1}}%
      \expandafter\def\csname LT5\endcsname{\color[rgb]{1,1,0}}%
      \expandafter\def\csname LT6\endcsname{\color[rgb]{0,0,0}}%
      \expandafter\def\csname LT7\endcsname{\color[rgb]{1,0.3,0}}%
      \expandafter\def\csname LT8\endcsname{\color[rgb]{0.5,0.5,0.5}}%
    \else
      \def\colorrgb#1{\color{black}}%
      \def\colorgray#1{\color[gray]{#1}}%
      \expandafter\def\csname LTw\endcsname{\color{white}}%
      \expandafter\def\csname LTb\endcsname{\color{black}}%
      \expandafter\def\csname LTa\endcsname{\color{black}}%
      \expandafter\def\csname LT0\endcsname{\color{black}}%
      \expandafter\def\csname LT1\endcsname{\color{black}}%
      \expandafter\def\csname LT2\endcsname{\color{black}}%
      \expandafter\def\csname LT3\endcsname{\color{black}}%
      \expandafter\def\csname LT4\endcsname{\color{black}}%
      \expandafter\def\csname LT5\endcsname{\color{black}}%
      \expandafter\def\csname LT6\endcsname{\color{black}}%
      \expandafter\def\csname LT7\endcsname{\color{black}}%
      \expandafter\def\csname LT8\endcsname{\color{black}}%
    \fi
  \fi
    \setlength{\unitlength}{0.0500bp}%
    \ifx\gptboxheight\undefined%
      \newlength{\gptboxheight}%
      \newlength{\gptboxwidth}%
      \newsavebox{\gptboxtext}%
    \fi%
    \setlength{\fboxrule}{0.5pt}%
    \setlength{\fboxsep}{1pt}%
\begin{picture}(7200.00,4320.00)%
    \gplgaddtomacro\gplbacktext{%
      \colorrgb{0.50,0.50,0.50}%
      \put(747,595){\makebox(0,0)[r]{\strut{}$-0.1$}}%
      \colorrgb{0.50,0.50,0.50}%
      \put(747,986){\makebox(0,0)[r]{\strut{}$0$}}%
      \colorrgb{0.50,0.50,0.50}%
      \put(747,1377){\makebox(0,0)[r]{\strut{}$0.1$}}%
      \colorrgb{0.50,0.50,0.50}%
      \put(747,1768){\makebox(0,0)[r]{\strut{}$0.2$}}%
      \colorrgb{0.50,0.50,0.50}%
      \put(747,2159){\makebox(0,0)[r]{\strut{}$0.3$}}%
      \colorrgb{0.50,0.50,0.50}%
      \put(747,2551){\makebox(0,0)[r]{\strut{}$0.4$}}%
      \colorrgb{0.50,0.50,0.50}%
      \put(747,2942){\makebox(0,0)[r]{\strut{}$0.5$}}%
      \colorrgb{0.50,0.50,0.50}%
      \put(747,3333){\makebox(0,0)[r]{\strut{}$0.6$}}%
      \colorrgb{0.50,0.50,0.50}%
      \put(747,3724){\makebox(0,0)[r]{\strut{}$0.7$}}%
      \colorrgb{0.50,0.50,0.50}%
      \put(747,4115){\makebox(0,0)[r]{\strut{}$0.8$}}%
      \colorrgb{0.50,0.50,0.50}%
      \put(911,409){\makebox(0,0){\strut{}$0$}}%
      \colorrgb{0.50,0.50,0.50}%
      \put(1546,409){\makebox(0,0){\strut{}$0.01$}}%
      \colorrgb{0.50,0.50,0.50}%
      \put(2182,409){\makebox(0,0){\strut{}$0.02$}}%
      \colorrgb{0.50,0.50,0.50}%
      \put(2818,409){\makebox(0,0){\strut{}$0.03$}}%
      \colorrgb{0.50,0.50,0.50}%
      \put(3454,409){\makebox(0,0){\strut{}$0.04$}}%
      \colorrgb{0.50,0.50,0.50}%
      \put(4090,409){\makebox(0,0){\strut{}$0.05$}}%
      \colorrgb{0.50,0.50,0.50}%
      \put(4726,409){\makebox(0,0){\strut{}$0.06$}}%
      \colorrgb{0.50,0.50,0.50}%
      \put(5361,409){\makebox(0,0){\strut{}$0.07$}}%
      \colorrgb{0.50,0.50,0.50}%
      \put(5997,409){\makebox(0,0){\strut{}$0.08$}}%
      \colorrgb{0.50,0.50,0.50}%
      \put(6633,409){\makebox(0,0){\strut{}$0.09$}}%
      \csname LTb\endcsname%
      \put(4789,1416){\makebox(0,0)[l]{\strut{}$a\mu = 0.04$}}%
    }%
    \gplgaddtomacro\gplfronttext{%
      \csname LTb\endcsname%
      \put(144,2355){\rotatebox{-270}{\makebox(0,0){\strut{}$\ja  \times 10^{ 3}\;[a^{-3}]$}}}%
      \csname LTb\endcsname%
      \put(4024,130){\makebox(0,0){\strut{}$qB\;[a^{-2}]$}}%
      \csname LTb\endcsname%
      \put(2787,3948){\makebox(0,0)[r]{\strut{}$95.45\%$ CI $40$ Est.}}%
      \csname LTb\endcsname%
      \put(2787,3762){\makebox(0,0)[r]{\strut{}$95.45\%$ CI $60$ Est.}}%
      \csname LTb\endcsname%
      \put(2787,3576){\makebox(0,0)[r]{\strut{}$95.45\%$ CI $80$ Est.}}%
      \csname LTb\endcsname%
      \put(2787,3390){\makebox(0,0)[r]{\strut{}$\scscf$}}%
    }%
    \gplbacktext
    \put(0,0){\includegraphics{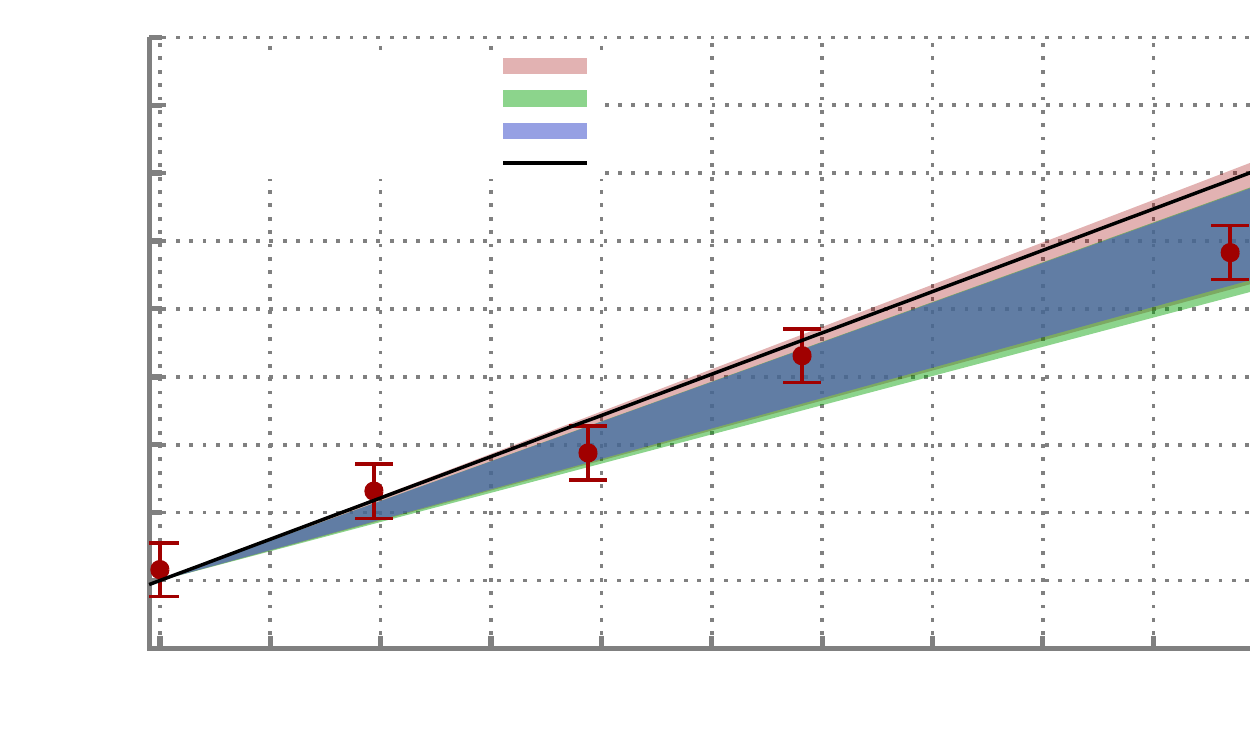}}%
    \gplfronttext
  \end{picture}%
\endgroup

%% file: CI_graph
\begingroup
  \makeatletter
  \providecommand\color[2][]{%
    \GenericError{(gnuplot) \space\space\space\@spaces}{%
      Package color not loaded in conjunction with
      terminal option `colourtext'%
    }{See the gnuplot documentation for explanation.%
    }{Either use 'blacktext' in gnuplot or load the package
      color.sty in LaTeX.}%
    \renewcommand\color[2][]{}%
  }%
  \providecommand\includegraphics[2][]{%
    \GenericError{(gnuplot) \space\space\space\@spaces}{%
      Package graphicx or graphics not loaded%
    }{See the gnuplot documentation for explanation.%
    }{The gnuplot epslatex terminal needs graphicx.sty or graphics.sty.}%
    \renewcommand\includegraphics[2][]{}%
  }%
  \providecommand\rotatebox[2]{#2}%
  \@ifundefined{ifGPcolor}{%
    \newif\ifGPcolor
    \GPcolortrue
  }{}%
  \@ifundefined{ifGPblacktext}{%
    \newif\ifGPblacktext
    \GPblacktextfalse
  }{}%
  \let\gplgaddtomacro\g@addto@macro
  \gdef\gplbacktext{}%
  \gdef\gplfronttext{}%
  \makeatother
  \ifGPblacktext
    \def\colorrgb#1{}%
    \def\colorgray#1{}%
  \else
    \ifGPcolor
      \def\colorrgb#1{\color[rgb]{#1}}%
      \def\colorgray#1{\color[gray]{#1}}%
      \expandafter\def\csname LTw\endcsname{\color{white}}%
      \expandafter\def\csname LTb\endcsname{\color{black}}%
      \expandafter\def\csname LTa\endcsname{\color{black}}%
      \expandafter\def\csname LT0\endcsname{\color[rgb]{1,0,0}}%
      \expandafter\def\csname LT1\endcsname{\color[rgb]{0,1,0}}%
      \expandafter\def\csname LT2\endcsname{\color[rgb]{0,0,1}}%
      \expandafter\def\csname LT3\endcsname{\color[rgb]{1,0,1}}%
      \expandafter\def\csname LT4\endcsname{\color[rgb]{0,1,1}}%
      \expandafter\def\csname LT5\endcsname{\color[rgb]{1,1,0}}%
      \expandafter\def\csname LT6\endcsname{\color[rgb]{0,0,0}}%
      \expandafter\def\csname LT7\endcsname{\color[rgb]{1,0.3,0}}%
      \expandafter\def\csname LT8\endcsname{\color[rgb]{0.5,0.5,0.5}}%
    \else
      \def\colorrgb#1{\color{black}}%
      \def\colorgray#1{\color[gray]{#1}}%
      \expandafter\def\csname LTw\endcsname{\color{white}}%
      \expandafter\def\csname LTb\endcsname{\color{black}}%
      \expandafter\def\csname LTa\endcsname{\color{black}}%
      \expandafter\def\csname LT0\endcsname{\color{black}}%
      \expandafter\def\csname LT1\endcsname{\color{black}}%
      \expandafter\def\csname LT2\endcsname{\color{black}}%
      \expandafter\def\csname LT3\endcsname{\color{black}}%
      \expandafter\def\csname LT4\endcsname{\color{black}}%
      \expandafter\def\csname LT5\endcsname{\color{black}}%
      \expandafter\def\csname LT6\endcsname{\color{black}}%
      \expandafter\def\csname LT7\endcsname{\color{black}}%
      \expandafter\def\csname LT8\endcsname{\color{black}}%
    \fi
  \fi
    \setlength{\unitlength}{0.0500bp}%
    \ifx\gptboxheight\undefined%
      \newlength{\gptboxheight}%
      \newlength{\gptboxwidth}%
      \newsavebox{\gptboxtext}%
    \fi%
    \setlength{\fboxrule}{0.5pt}%
    \setlength{\fboxsep}{1pt}%
\begin{picture}(7200.00,4320.00)%
    \gplgaddtomacro\gplbacktext{%
      \colorrgb{0.50,0.50,0.50}%
      \put(543,595){\makebox(0,0)[r]{\strut{}$0$}}%
      \colorrgb{0.50,0.50,0.50}%
      \put(543,1475){\makebox(0,0)[r]{\strut{}$0.5$}}%
      \colorrgb{0.50,0.50,0.50}%
      \put(543,2355){\makebox(0,0)[r]{\strut{}$1$}}%
      \colorrgb{0.50,0.50,0.50}%
      \put(543,3235){\makebox(0,0)[r]{\strut{}$1.5$}}%
      \colorrgb{0.50,0.50,0.50}%
      \put(543,4115){\makebox(0,0)[r]{\strut{}$2$}}%
      \colorrgb{0.50,0.50,0.50}%
      \put(1120,409){\makebox(0,0){\strut{}$0.040$}}%
      \colorrgb{0.50,0.50,0.50}%
      \put(2070,409){\makebox(0,0){\strut{}$0.230$}}%
      \colorrgb{0.50,0.50,0.50}%
      \put(3020,409){\makebox(0,0){\strut{}$0.040$}}%
      \colorrgb{0.50,0.50,0.50}%
      \put(3969,409){\makebox(0,0){\strut{}$0.050$}}%
      \colorrgb{0.50,0.50,0.50}%
      \put(4919,409){\makebox(0,0){\strut{}$0.300$}}%
      \colorrgb{0.50,0.50,0.50}%
      \put(5869,409){\makebox(0,0){\strut{}$0.300$}}%
      \colorrgb{0.50,0.50,0.50}%
      \put(6819,409){\makebox(0,0){\strut{}$0.300$}}%
      \csname LTb\endcsname%
      \put(1880,3587){\makebox(0,0)[l]{\strut{}$T > \Tc$}}%
      \csname LTb\endcsname%
      \put(5204,3587){\makebox(0,0)[l]{\strut{}$T < \Tc$}}%
      \csname LTb\endcsname%
      \put(2830,947){\makebox(0,0)[l]{\strut{}$\!|Q|\!=\!1$}}%
      \csname LTb\endcsname%
      \put(1880,947){\makebox(0,0)[l]{\strut{}$\!|Q|\!=\!0$}}%
      \csname LTb\endcsname%
      \put(930,947){\makebox(0,0)[l]{\strut{}$\!|Q|\!=\!0$}}%
      \csname LTb\endcsname%
      \put(3780,947){\makebox(0,0)[l]{\strut{}$\!|Q|\!=\!0$}}%
      \csname LTb\endcsname%
      \put(4729,947){\makebox(0,0)[l]{\strut{}$\!|Q|\!=\!0$}}%
      \csname LTb\endcsname%
      \put(5679,947){\makebox(0,0)[l]{\strut{}$\!|Q|\!=\!1$}}%
      \csname LTb\endcsname%
      \put(6629,947){\makebox(0,0)[l]{\strut{}$\!|Q|\!=\!2$}}%
    }%
    \gplgaddtomacro\gplfronttext{%
      \csname LTb\endcsname%
      \put(144,2355){\rotatebox{-270}{\makebox(0,0){\strut{}$\scsc/\scscf$}}}%
      \csname LTb\endcsname%
      \put(3922,130){\makebox(0,0){\strut{}$a\mu$}}%
    }%
    \gplbacktext
    \put(0,0){\includegraphics{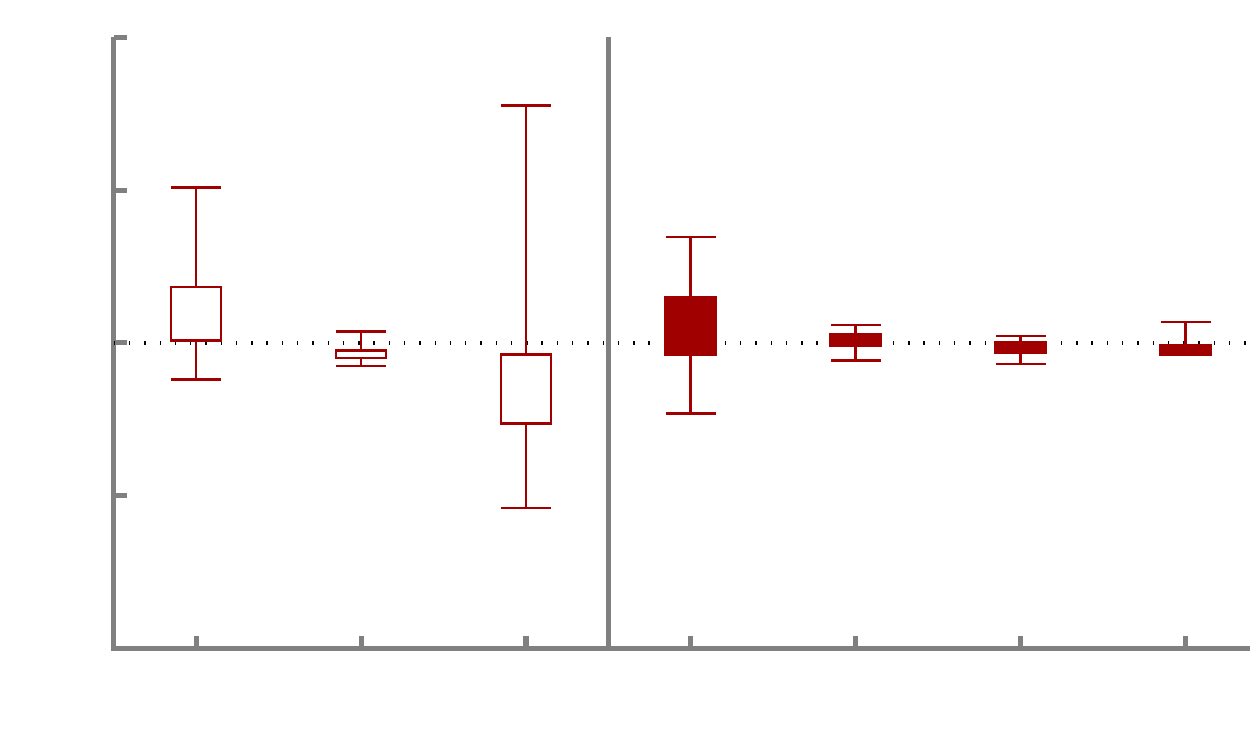}}%
    \gplfronttext
  \end{picture}%
\endgroup

%% file: 8x8b81_mu0100_all_mo
\begingroup
  \makeatletter
  \providecommand\color[2][]{%
    \GenericError{(gnuplot) \space\space\space\@spaces}{%
      Package color not loaded in conjunction with
      terminal option `colourtext'%
    }{See the gnuplot documentation for explanation.%
    }{Either use 'blacktext' in gnuplot or load the package
      color.sty in LaTeX.}%
    \renewcommand\color[2][]{}%
  }%
  \providecommand\includegraphics[2][]{%
    \GenericError{(gnuplot) \space\space\space\@spaces}{%
      Package graphicx or graphics not loaded%
    }{See the gnuplot documentation for explanation.%
    }{The gnuplot epslatex terminal needs graphicx.sty or graphics.sty.}%
    \renewcommand\includegraphics[2][]{}%
  }%
  \providecommand\rotatebox[2]{#2}%
  \@ifundefined{ifGPcolor}{%
    \newif\ifGPcolor
    \GPcolortrue
  }{}%
  \@ifundefined{ifGPblacktext}{%
    \newif\ifGPblacktext
    \GPblacktextfalse
  }{}%
  \let\gplgaddtomacro\g@addto@macro
  \gdef\gplbacktext{}%
  \gdef\gplfronttext{}%
  \makeatother
  \ifGPblacktext
    \def\colorrgb#1{}%
    \def\colorgray#1{}%
  \else
    \ifGPcolor
      \def\colorrgb#1{\color[rgb]{#1}}%
      \def\colorgray#1{\color[gray]{#1}}%
      \expandafter\def\csname LTw\endcsname{\color{white}}%
      \expandafter\def\csname LTb\endcsname{\color{black}}%
      \expandafter\def\csname LTa\endcsname{\color{black}}%
      \expandafter\def\csname LT0\endcsname{\color[rgb]{1,0,0}}%
      \expandafter\def\csname LT1\endcsname{\color[rgb]{0,1,0}}%
      \expandafter\def\csname LT2\endcsname{\color[rgb]{0,0,1}}%
      \expandafter\def\csname LT3\endcsname{\color[rgb]{1,0,1}}%
      \expandafter\def\csname LT4\endcsname{\color[rgb]{0,1,1}}%
      \expandafter\def\csname LT5\endcsname{\color[rgb]{1,1,0}}%
      \expandafter\def\csname LT6\endcsname{\color[rgb]{0,0,0}}%
      \expandafter\def\csname LT7\endcsname{\color[rgb]{1,0.3,0}}%
      \expandafter\def\csname LT8\endcsname{\color[rgb]{0.5,0.5,0.5}}%
    \else
      \def\colorrgb#1{\color{black}}%
      \def\colorgray#1{\color[gray]{#1}}%
      \expandafter\def\csname LTw\endcsname{\color{white}}%
      \expandafter\def\csname LTb\endcsname{\color{black}}%
      \expandafter\def\csname LTa\endcsname{\color{black}}%
      \expandafter\def\csname LT0\endcsname{\color{black}}%
      \expandafter\def\csname LT1\endcsname{\color{black}}%
      \expandafter\def\csname LT2\endcsname{\color{black}}%
      \expandafter\def\csname LT3\endcsname{\color{black}}%
      \expandafter\def\csname LT4\endcsname{\color{black}}%
      \expandafter\def\csname LT5\endcsname{\color{black}}%
      \expandafter\def\csname LT6\endcsname{\color{black}}%
      \expandafter\def\csname LT7\endcsname{\color{black}}%
      \expandafter\def\csname LT8\endcsname{\color{black}}%
    \fi
  \fi
    \setlength{\unitlength}{0.0500bp}%
    \ifx\gptboxheight\undefined%
      \newlength{\gptboxheight}%
      \newlength{\gptboxwidth}%
      \newsavebox{\gptboxtext}%
    \fi%
    \setlength{\fboxrule}{0.5pt}%
    \setlength{\fboxsep}{1pt}%
\begin{picture}(7200.00,4320.00)%
    \gplgaddtomacro\gplbacktext{%
      \colorrgb{0.50,0.50,0.50}%
      \put(441,681){\makebox(0,0)[r]{\strut{}$0$}}%
      \colorrgb{0.50,0.50,0.50}%
      \put(441,1539){\makebox(0,0)[r]{\strut{}$2$}}%
      \colorrgb{0.50,0.50,0.50}%
      \put(441,2398){\makebox(0,0)[r]{\strut{}$4$}}%
      \colorrgb{0.50,0.50,0.50}%
      \put(441,3256){\makebox(0,0)[r]{\strut{}$6$}}%
      \colorrgb{0.50,0.50,0.50}%
      \put(441,4115){\makebox(0,0)[r]{\strut{}$8$}}%
      \colorrgb{0.50,0.50,0.50}%
      \put(543,409){\makebox(0,0){\strut{}$0$}}%
      \colorrgb{0.50,0.50,0.50}%
      \put(1221,409){\makebox(0,0){\strut{}$0.05$}}%
      \colorrgb{0.50,0.50,0.50}%
      \put(2577,409){\makebox(0,0){\strut{}$0.15$}}%
      \colorrgb{0.50,0.50,0.50}%
      \put(3933,409){\makebox(0,0){\strut{}$0.25$}}%
      \colorrgb{0.50,0.50,0.50}%
      \put(5289,409){\makebox(0,0){\strut{}$0.35$}}%
      \colorrgb{0.50,0.50,0.50}%
      \put(6645,409){\makebox(0,0){\strut{}$0.45$}}%
      \csname LTb\endcsname%
      \put(4272,3686){\makebox(0,0)[l]{\strut{}$ T < \Tc$}}%
    }%
    \gplgaddtomacro\gplfronttext{%
      \csname LTb\endcsname%
      \put(144,2355){\rotatebox{-270}{\makebox(0,0){\strut{}$\ja  \times 10^{ 3}\;[a^{-3}]$}}}%
      \csname LTb\endcsname%
      \put(3871,130){\makebox(0,0){\strut{}$qB\;[a^{-2}]$}}%
      \csname LTb\endcsname%
      \put(1331,3948){\makebox(0,0)[l]{\strut{}$|Q|=00$}}%
      \csname LTb\endcsname%
      \put(1331,3762){\makebox(0,0)[l]{\strut{}$|Q|=01$}}%
      \csname LTb\endcsname%
      \put(1331,3576){\makebox(0,0)[l]{\strut{}$|Q|=02$}}%
      \csname LTb\endcsname%
      \put(1331,3390){\makebox(0,0)[l]{\strut{}$|Q|=00 , \mq=0.0$}}%
    }%
    \gplbacktext
    \put(0,0){\includegraphics{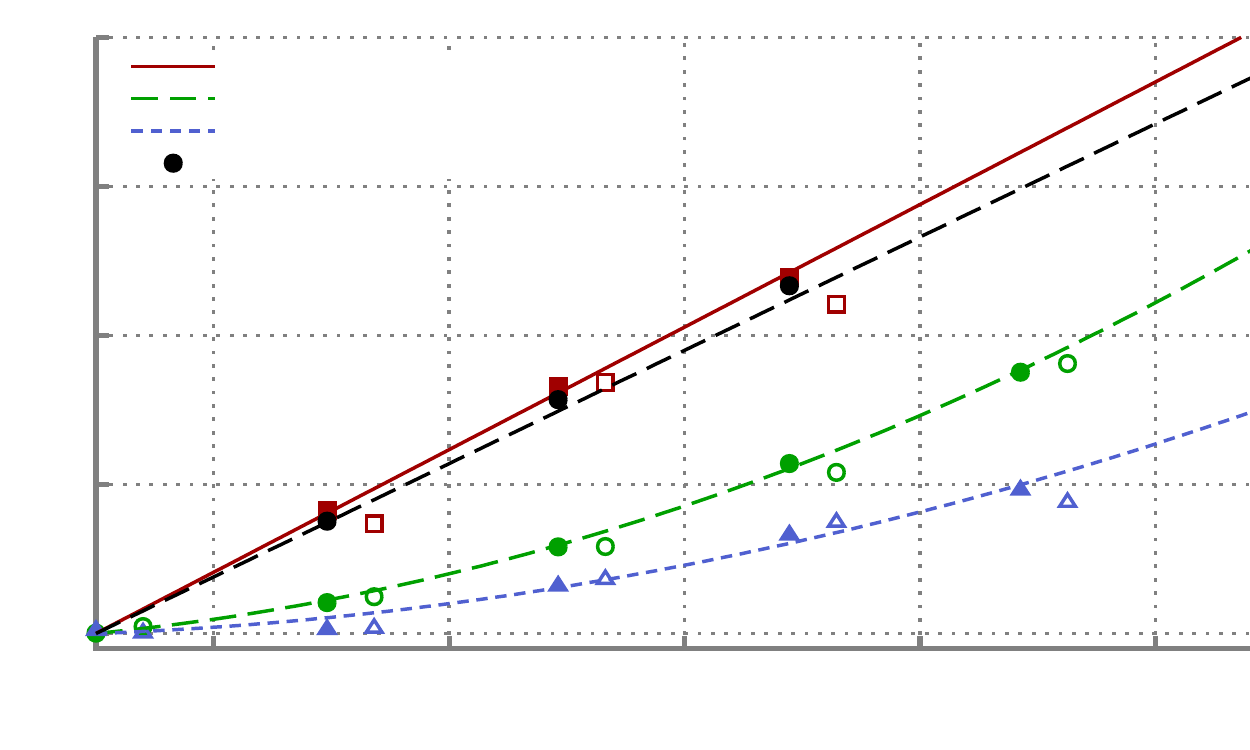}}%
    \gplfronttext
  \end{picture}%
\endgroup

%% file: 14x14b81_slope_CI
\begingroup
  \makeatletter
  \providecommand\color[2][]{%
    \GenericError{(gnuplot) \space\space\space\@spaces}{%
      Package color not loaded in conjunction with
      terminal option `colourtext'%
    }{See the gnuplot documentation for explanation.%
    }{Either use 'blacktext' in gnuplot or load the package
      color.sty in LaTeX.}%
    \renewcommand\color[2][]{}%
  }%
  \providecommand\includegraphics[2][]{%
    \GenericError{(gnuplot) \space\space\space\@spaces}{%
      Package graphicx or graphics not loaded%
    }{See the gnuplot documentation for explanation.%
    }{The gnuplot epslatex terminal needs graphicx.sty or graphics.sty.}%
    \renewcommand\includegraphics[2][]{}%
  }%
  \providecommand\rotatebox[2]{#2}%
  \@ifundefined{ifGPcolor}{%
    \newif\ifGPcolor
    \GPcolortrue
  }{}%
  \@ifundefined{ifGPblacktext}{%
    \newif\ifGPblacktext
    \GPblacktextfalse
  }{}%
  \let\gplgaddtomacro\g@addto@macro
  \gdef\gplbacktext{}%
  \gdef\gplfronttext{}%
  \makeatother
  \ifGPblacktext
    \def\colorrgb#1{}%
    \def\colorgray#1{}%
  \else
    \ifGPcolor
      \def\colorrgb#1{\color[rgb]{#1}}%
      \def\colorgray#1{\color[gray]{#1}}%
      \expandafter\def\csname LTw\endcsname{\color{white}}%
      \expandafter\def\csname LTb\endcsname{\color{black}}%
      \expandafter\def\csname LTa\endcsname{\color{black}}%
      \expandafter\def\csname LT0\endcsname{\color[rgb]{1,0,0}}%
      \expandafter\def\csname LT1\endcsname{\color[rgb]{0,1,0}}%
      \expandafter\def\csname LT2\endcsname{\color[rgb]{0,0,1}}%
      \expandafter\def\csname LT3\endcsname{\color[rgb]{1,0,1}}%
      \expandafter\def\csname LT4\endcsname{\color[rgb]{0,1,1}}%
      \expandafter\def\csname LT5\endcsname{\color[rgb]{1,1,0}}%
      \expandafter\def\csname LT6\endcsname{\color[rgb]{0,0,0}}%
      \expandafter\def\csname LT7\endcsname{\color[rgb]{1,0.3,0}}%
      \expandafter\def\csname LT8\endcsname{\color[rgb]{0.5,0.5,0.5}}%
    \else
      \def\colorrgb#1{\color{black}}%
      \def\colorgray#1{\color[gray]{#1}}%
      \expandafter\def\csname LTw\endcsname{\color{white}}%
      \expandafter\def\csname LTb\endcsname{\color{black}}%
      \expandafter\def\csname LTa\endcsname{\color{black}}%
      \expandafter\def\csname LT0\endcsname{\color{black}}%
      \expandafter\def\csname LT1\endcsname{\color{black}}%
      \expandafter\def\csname LT2\endcsname{\color{black}}%
      \expandafter\def\csname LT3\endcsname{\color{black}}%
      \expandafter\def\csname LT4\endcsname{\color{black}}%
      \expandafter\def\csname LT5\endcsname{\color{black}}%
      \expandafter\def\csname LT6\endcsname{\color{black}}%
      \expandafter\def\csname LT7\endcsname{\color{black}}%
      \expandafter\def\csname LT8\endcsname{\color{black}}%
    \fi
  \fi
    \setlength{\unitlength}{0.0500bp}%
    \ifx\gptboxheight\undefined%
      \newlength{\gptboxheight}%
      \newlength{\gptboxwidth}%
      \newsavebox{\gptboxtext}%
    \fi%
    \setlength{\fboxrule}{0.5pt}%
    \setlength{\fboxsep}{1pt}%
\begin{picture}(7200.00,4320.00)%
    \gplgaddtomacro\gplbacktext{%
      \colorrgb{0.50,0.50,0.50}%
      \put(543,595){\makebox(0,0)[r]{\strut{}$0$}}%
      \colorrgb{0.50,0.50,0.50}%
      \put(543,986){\makebox(0,0)[r]{\strut{}$2$}}%
      \colorrgb{0.50,0.50,0.50}%
      \put(543,1377){\makebox(0,0)[r]{\strut{}$4$}}%
      \colorrgb{0.50,0.50,0.50}%
      \put(543,1768){\makebox(0,0)[r]{\strut{}$6$}}%
      \colorrgb{0.50,0.50,0.50}%
      \put(543,2159){\makebox(0,0)[r]{\strut{}$8$}}%
      \colorrgb{0.50,0.50,0.50}%
      \put(543,2551){\makebox(0,0)[r]{\strut{}$10$}}%
      \colorrgb{0.50,0.50,0.50}%
      \put(543,2942){\makebox(0,0)[r]{\strut{}$12$}}%
      \colorrgb{0.50,0.50,0.50}%
      \put(543,3333){\makebox(0,0)[r]{\strut{}$14$}}%
      \colorrgb{0.50,0.50,0.50}%
      \put(543,3724){\makebox(0,0)[r]{\strut{}$16$}}%
      \colorrgb{0.50,0.50,0.50}%
      \put(543,4115){\makebox(0,0)[r]{\strut{}$18$}}%
      \colorrgb{0.50,0.50,0.50}%
      \put(645,409){\makebox(0,0){\strut{}$0$}}%
      \colorrgb{0.50,0.50,0.50}%
      \put(1570,409){\makebox(0,0){\strut{}$0.05$}}%
      \colorrgb{0.50,0.50,0.50}%
      \put(2495,409){\makebox(0,0){\strut{}$0.1$}}%
      \colorrgb{0.50,0.50,0.50}%
      \put(3420,409){\makebox(0,0){\strut{}$0.15$}}%
      \colorrgb{0.50,0.50,0.50}%
      \put(4345,409){\makebox(0,0){\strut{}$0.2$}}%
      \colorrgb{0.50,0.50,0.50}%
      \put(5271,409){\makebox(0,0){\strut{}$0.25$}}%
      \colorrgb{0.50,0.50,0.50}%
      \put(6196,409){\makebox(0,0){\strut{}$0.3$}}%
      \csname LTb\endcsname%
      \put(4364,1213){\makebox(0,0)[l]{\strut{}\scriptsize  $a\mu = 0.05 $}}%
      \csname LTb\endcsname%
      \put(4364,3172){\makebox(0,0)[l]{\strut{}\scriptsize  $a\mu = 0.30 $}}%
    }%
    \gplgaddtomacro\gplfronttext{%
      \csname LTb\endcsname%
      \put(144,2355){\rotatebox{-270}{\makebox(0,0){\strut{}$\ja  \times 10^{ 3}\;[a^{-3}]$}}}%
      \csname LTb\endcsname%
      \put(3922,130){\makebox(0,0){\strut{}$qB\;[a^{-2}]$}}%
      \csname LTb\endcsname%
      \put(2277,3948){\makebox(0,0)[r]{\strut{}$95.45\%$ CI $40$ Est.}}%
      \csname LTb\endcsname%
      \put(2277,3762){\makebox(0,0)[r]{\strut{}$95.45\%$ CI $60$ Est.}}%
      \csname LTb\endcsname%
      \put(2277,3576){\makebox(0,0)[r]{\strut{}$95.45\%$ CI $80$ Est.}}%
      \csname LTb\endcsname%
      \put(2277,3390){\makebox(0,0)[r]{\strut{}$\scscf$}}%
    }%
    \gplbacktext
    \put(0,0){\includegraphics{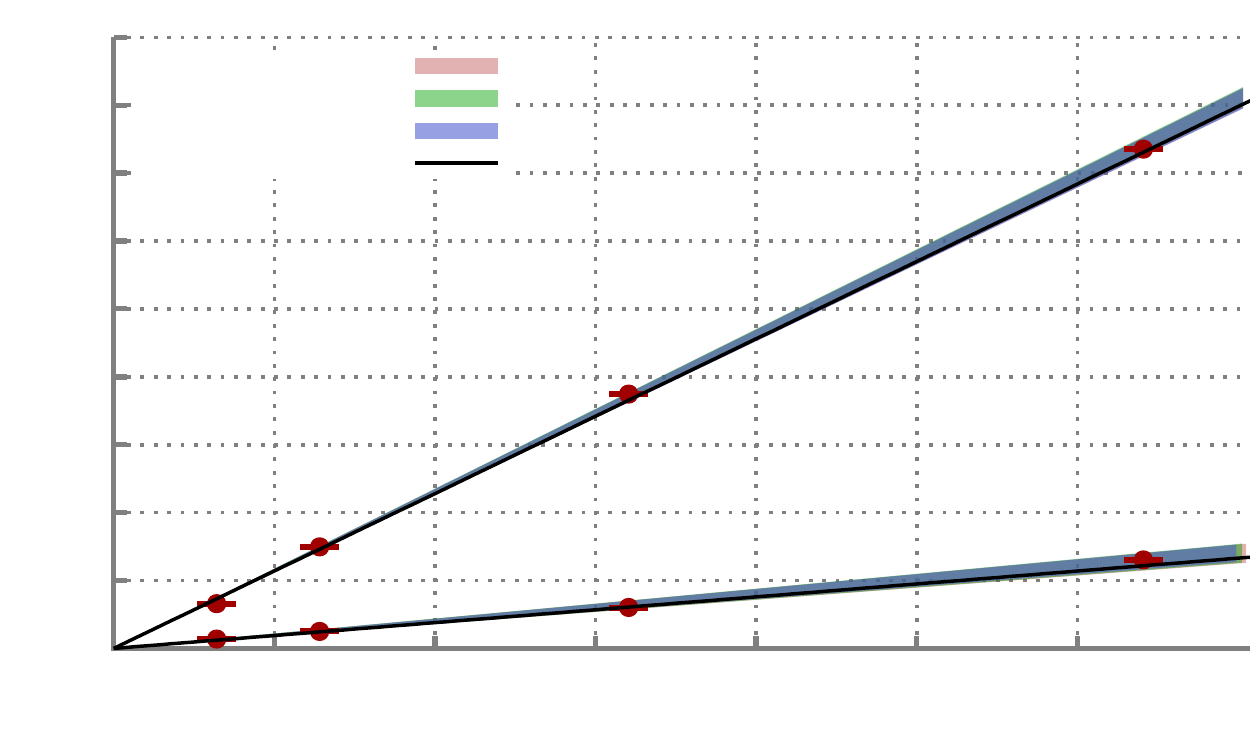}}%
    \gplfronttext
  \end{picture}%
\endgroup

%% file: 14x14b81_mu0300_slope_CI_Q01
\begingroup
  \makeatletter
  \providecommand\color[2][]{%
    \GenericError{(gnuplot) \space\space\space\@spaces}{%
      Package color not loaded in conjunction with
      terminal option `colourtext'%
    }{See the gnuplot documentation for explanation.%
    }{Either use 'blacktext' in gnuplot or load the package
      color.sty in LaTeX.}%
    \renewcommand\color[2][]{}%
  }%
  \providecommand\includegraphics[2][]{%
    \GenericError{(gnuplot) \space\space\space\@spaces}{%
      Package graphicx or graphics not loaded%
    }{See the gnuplot documentation for explanation.%
    }{The gnuplot epslatex terminal needs graphicx.sty or graphics.sty.}%
    \renewcommand\includegraphics[2][]{}%
  }%
  \providecommand\rotatebox[2]{#2}%
  \@ifundefined{ifGPcolor}{%
    \newif\ifGPcolor
    \GPcolortrue
  }{}%
  \@ifundefined{ifGPblacktext}{%
    \newif\ifGPblacktext
    \GPblacktextfalse
  }{}%
  \let\gplgaddtomacro\g@addto@macro
  \gdef\gplbacktext{}%
  \gdef\gplfronttext{}%
  \makeatother
  \ifGPblacktext
    \def\colorrgb#1{}%
    \def\colorgray#1{}%
  \else
    \ifGPcolor
      \def\colorrgb#1{\color[rgb]{#1}}%
      \def\colorgray#1{\color[gray]{#1}}%
      \expandafter\def\csname LTw\endcsname{\color{white}}%
      \expandafter\def\csname LTb\endcsname{\color{black}}%
      \expandafter\def\csname LTa\endcsname{\color{black}}%
      \expandafter\def\csname LT0\endcsname{\color[rgb]{1,0,0}}%
      \expandafter\def\csname LT1\endcsname{\color[rgb]{0,1,0}}%
      \expandafter\def\csname LT2\endcsname{\color[rgb]{0,0,1}}%
      \expandafter\def\csname LT3\endcsname{\color[rgb]{1,0,1}}%
      \expandafter\def\csname LT4\endcsname{\color[rgb]{0,1,1}}%
      \expandafter\def\csname LT5\endcsname{\color[rgb]{1,1,0}}%
      \expandafter\def\csname LT6\endcsname{\color[rgb]{0,0,0}}%
      \expandafter\def\csname LT7\endcsname{\color[rgb]{1,0.3,0}}%
      \expandafter\def\csname LT8\endcsname{\color[rgb]{0.5,0.5,0.5}}%
    \else
      \def\colorrgb#1{\color{black}}%
      \def\colorgray#1{\color[gray]{#1}}%
      \expandafter\def\csname LTw\endcsname{\color{white}}%
      \expandafter\def\csname LTb\endcsname{\color{black}}%
      \expandafter\def\csname LTa\endcsname{\color{black}}%
      \expandafter\def\csname LT0\endcsname{\color{black}}%
      \expandafter\def\csname LT1\endcsname{\color{black}}%
      \expandafter\def\csname LT2\endcsname{\color{black}}%
      \expandafter\def\csname LT3\endcsname{\color{black}}%
      \expandafter\def\csname LT4\endcsname{\color{black}}%
      \expandafter\def\csname LT5\endcsname{\color{black}}%
      \expandafter\def\csname LT6\endcsname{\color{black}}%
      \expandafter\def\csname LT7\endcsname{\color{black}}%
      \expandafter\def\csname LT8\endcsname{\color{black}}%
    \fi
  \fi
    \setlength{\unitlength}{0.0500bp}%
    \ifx\gptboxheight\undefined%
      \newlength{\gptboxheight}%
      \newlength{\gptboxwidth}%
      \newsavebox{\gptboxtext}%
    \fi%
    \setlength{\fboxrule}{0.5pt}%
    \setlength{\fboxsep}{1pt}%
\begin{picture}(7200.00,4320.00)%
    \gplgaddtomacro\gplbacktext{%
      \colorrgb{0.50,0.50,0.50}%
      \put(543,595){\makebox(0,0)[r]{\strut{}$0$}}%
      \colorrgb{0.50,0.50,0.50}%
      \put(543,986){\makebox(0,0)[r]{\strut{}$2$}}%
      \colorrgb{0.50,0.50,0.50}%
      \put(543,1377){\makebox(0,0)[r]{\strut{}$4$}}%
      \colorrgb{0.50,0.50,0.50}%
      \put(543,1768){\makebox(0,0)[r]{\strut{}$6$}}%
      \colorrgb{0.50,0.50,0.50}%
      \put(543,2159){\makebox(0,0)[r]{\strut{}$8$}}%
      \colorrgb{0.50,0.50,0.50}%
      \put(543,2551){\makebox(0,0)[r]{\strut{}$10$}}%
      \colorrgb{0.50,0.50,0.50}%
      \put(543,2942){\makebox(0,0)[r]{\strut{}$12$}}%
      \colorrgb{0.50,0.50,0.50}%
      \put(543,3333){\makebox(0,0)[r]{\strut{}$14$}}%
      \colorrgb{0.50,0.50,0.50}%
      \put(543,3724){\makebox(0,0)[r]{\strut{}$16$}}%
      \colorrgb{0.50,0.50,0.50}%
      \put(543,4115){\makebox(0,0)[r]{\strut{}$18$}}%
      \colorrgb{0.50,0.50,0.50}%
      \put(645,409){\makebox(0,0){\strut{}$0$}}%
      \colorrgb{0.50,0.50,0.50}%
      \put(1570,409){\makebox(0,0){\strut{}$0.05$}}%
      \colorrgb{0.50,0.50,0.50}%
      \put(2495,409){\makebox(0,0){\strut{}$0.1$}}%
      \colorrgb{0.50,0.50,0.50}%
      \put(3420,409){\makebox(0,0){\strut{}$0.15$}}%
      \colorrgb{0.50,0.50,0.50}%
      \put(4345,409){\makebox(0,0){\strut{}$0.2$}}%
      \colorrgb{0.50,0.50,0.50}%
      \put(5271,409){\makebox(0,0){\strut{}$0.25$}}%
      \colorrgb{0.50,0.50,0.50}%
      \put(6196,409){\makebox(0,0){\strut{}$0.3$}}%
      \colorrgb{0.50,0.50,0.50}%
      \put(7121,409){\makebox(0,0){\strut{}$0.35$}}%
      \csname LTb\endcsname%
      \put(5548,3528){\makebox(0,0)[l]{\strut{}$\!|Q|\!=\!1$}}%
    }%
    \gplgaddtomacro\gplfronttext{%
      \csname LTb\endcsname%
      \put(144,2355){\rotatebox{-270}{\makebox(0,0){\strut{}$\ja  \times 10^{ 3}\;[a^{-3}]$}}}%
      \csname LTb\endcsname%
      \put(3922,130){\makebox(0,0){\strut{}$qB\;[a^{-2}]$}}%
      \csname LTb\endcsname%
      \put(2277,3948){\makebox(0,0)[r]{\strut{}$95.45\%$ CI $20$ Est.}}%
      \csname LTb\endcsname%
      \put(2277,3762){\makebox(0,0)[r]{\strut{}$\scscf$}}%
    }%
    \gplbacktext
    \put(0,0){\includegraphics{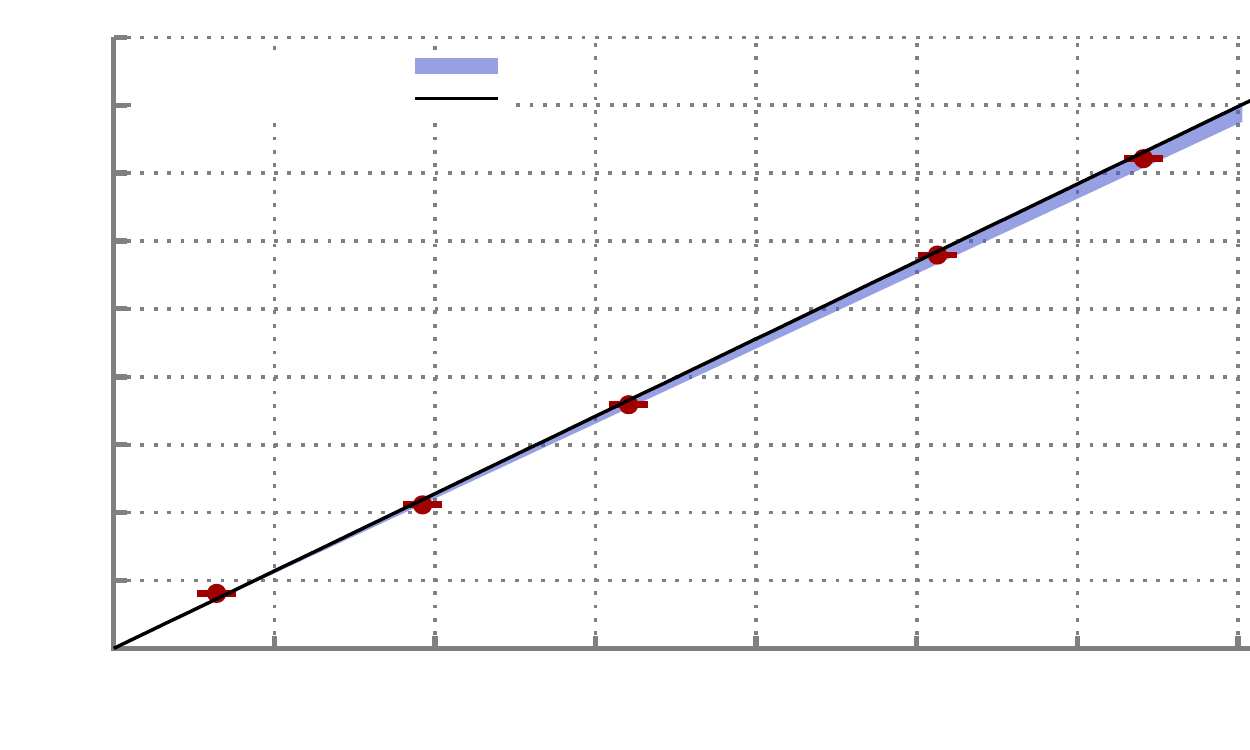}}%
    \gplfronttext
  \end{picture}%
\endgroup

%% file: 14x14b81_mu0300_slope_CI_Q02
\begingroup
  \makeatletter
  \providecommand\color[2][]{%
    \GenericError{(gnuplot) \space\space\space\@spaces}{%
      Package color not loaded in conjunction with
      terminal option `colourtext'%
    }{See the gnuplot documentation for explanation.%
    }{Either use 'blacktext' in gnuplot or load the package
      color.sty in LaTeX.}%
    \renewcommand\color[2][]{}%
  }%
  \providecommand\includegraphics[2][]{%
    \GenericError{(gnuplot) \space\space\space\@spaces}{%
      Package graphicx or graphics not loaded%
    }{See the gnuplot documentation for explanation.%
    }{The gnuplot epslatex terminal needs graphicx.sty or graphics.sty.}%
    \renewcommand\includegraphics[2][]{}%
  }%
  \providecommand\rotatebox[2]{#2}%
  \@ifundefined{ifGPcolor}{%
    \newif\ifGPcolor
    \GPcolortrue
  }{}%
  \@ifundefined{ifGPblacktext}{%
    \newif\ifGPblacktext
    \GPblacktextfalse
  }{}%
  \let\gplgaddtomacro\g@addto@macro
  \gdef\gplbacktext{}%
  \gdef\gplfronttext{}%
  \makeatother
  \ifGPblacktext
    \def\colorrgb#1{}%
    \def\colorgray#1{}%
  \else
    \ifGPcolor
      \def\colorrgb#1{\color[rgb]{#1}}%
      \def\colorgray#1{\color[gray]{#1}}%
      \expandafter\def\csname LTw\endcsname{\color{white}}%
      \expandafter\def\csname LTb\endcsname{\color{black}}%
      \expandafter\def\csname LTa\endcsname{\color{black}}%
      \expandafter\def\csname LT0\endcsname{\color[rgb]{1,0,0}}%
      \expandafter\def\csname LT1\endcsname{\color[rgb]{0,1,0}}%
      \expandafter\def\csname LT2\endcsname{\color[rgb]{0,0,1}}%
      \expandafter\def\csname LT3\endcsname{\color[rgb]{1,0,1}}%
      \expandafter\def\csname LT4\endcsname{\color[rgb]{0,1,1}}%
      \expandafter\def\csname LT5\endcsname{\color[rgb]{1,1,0}}%
      \expandafter\def\csname LT6\endcsname{\color[rgb]{0,0,0}}%
      \expandafter\def\csname LT7\endcsname{\color[rgb]{1,0.3,0}}%
      \expandafter\def\csname LT8\endcsname{\color[rgb]{0.5,0.5,0.5}}%
    \else
      \def\colorrgb#1{\color{black}}%
      \def\colorgray#1{\color[gray]{#1}}%
      \expandafter\def\csname LTw\endcsname{\color{white}}%
      \expandafter\def\csname LTb\endcsname{\color{black}}%
      \expandafter\def\csname LTa\endcsname{\color{black}}%
      \expandafter\def\csname LT0\endcsname{\color{black}}%
      \expandafter\def\csname LT1\endcsname{\color{black}}%
      \expandafter\def\csname LT2\endcsname{\color{black}}%
      \expandafter\def\csname LT3\endcsname{\color{black}}%
      \expandafter\def\csname LT4\endcsname{\color{black}}%
      \expandafter\def\csname LT5\endcsname{\color{black}}%
      \expandafter\def\csname LT6\endcsname{\color{black}}%
      \expandafter\def\csname LT7\endcsname{\color{black}}%
      \expandafter\def\csname LT8\endcsname{\color{black}}%
    \fi
  \fi
    \setlength{\unitlength}{0.0500bp}%
    \ifx\gptboxheight\undefined%
      \newlength{\gptboxheight}%
      \newlength{\gptboxwidth}%
      \newsavebox{\gptboxtext}%
    \fi%
    \setlength{\fboxrule}{0.5pt}%
    \setlength{\fboxsep}{1pt}%
\begin{picture}(7200.00,4320.00)%
    \gplgaddtomacro\gplbacktext{%
      \colorrgb{0.50,0.50,0.50}%
      \put(543,595){\makebox(0,0)[r]{\strut{}$0$}}%
      \colorrgb{0.50,0.50,0.50}%
      \put(543,986){\makebox(0,0)[r]{\strut{}$2$}}%
      \colorrgb{0.50,0.50,0.50}%
      \put(543,1377){\makebox(0,0)[r]{\strut{}$4$}}%
      \colorrgb{0.50,0.50,0.50}%
      \put(543,1768){\makebox(0,0)[r]{\strut{}$6$}}%
      \colorrgb{0.50,0.50,0.50}%
      \put(543,2159){\makebox(0,0)[r]{\strut{}$8$}}%
      \colorrgb{0.50,0.50,0.50}%
      \put(543,2551){\makebox(0,0)[r]{\strut{}$10$}}%
      \colorrgb{0.50,0.50,0.50}%
      \put(543,2942){\makebox(0,0)[r]{\strut{}$12$}}%
      \colorrgb{0.50,0.50,0.50}%
      \put(543,3333){\makebox(0,0)[r]{\strut{}$14$}}%
      \colorrgb{0.50,0.50,0.50}%
      \put(543,3724){\makebox(0,0)[r]{\strut{}$16$}}%
      \colorrgb{0.50,0.50,0.50}%
      \put(543,4115){\makebox(0,0)[r]{\strut{}$18$}}%
      \colorrgb{0.50,0.50,0.50}%
      \put(645,409){\makebox(0,0){\strut{}$0$}}%
      \colorrgb{0.50,0.50,0.50}%
      \put(1570,409){\makebox(0,0){\strut{}$0.05$}}%
      \colorrgb{0.50,0.50,0.50}%
      \put(2495,409){\makebox(0,0){\strut{}$0.1$}}%
      \colorrgb{0.50,0.50,0.50}%
      \put(3420,409){\makebox(0,0){\strut{}$0.15$}}%
      \colorrgb{0.50,0.50,0.50}%
      \put(4345,409){\makebox(0,0){\strut{}$0.2$}}%
      \colorrgb{0.50,0.50,0.50}%
      \put(5271,409){\makebox(0,0){\strut{}$0.25$}}%
      \colorrgb{0.50,0.50,0.50}%
      \put(6196,409){\makebox(0,0){\strut{}$0.3$}}%
      \colorrgb{0.50,0.50,0.50}%
      \put(7121,409){\makebox(0,0){\strut{}$0.35$}}%
      \csname LTb\endcsname%
      \put(5548,3528){\makebox(0,0)[l]{\strut{}$\!|Q|\!=\!2$}}%
    }%
    \gplgaddtomacro\gplfronttext{%
      \csname LTb\endcsname%
      \put(144,2355){\rotatebox{-270}{\makebox(0,0){\strut{}$\ja  \times 10^{ 3}\;[a^{-3}]$}}}%
      \csname LTb\endcsname%
      \put(3922,130){\makebox(0,0){\strut{}$qB\;[a^{-2}]$}}%
      \csname LTb\endcsname%
      \put(2277,3948){\makebox(0,0)[r]{\strut{}$95.45\%$ CI $20$ Est.}}%
      \csname LTb\endcsname%
      \put(2277,3762){\makebox(0,0)[r]{\strut{}$\scscf$}}%
    }%
    \gplbacktext
    \put(0,0){\includegraphics{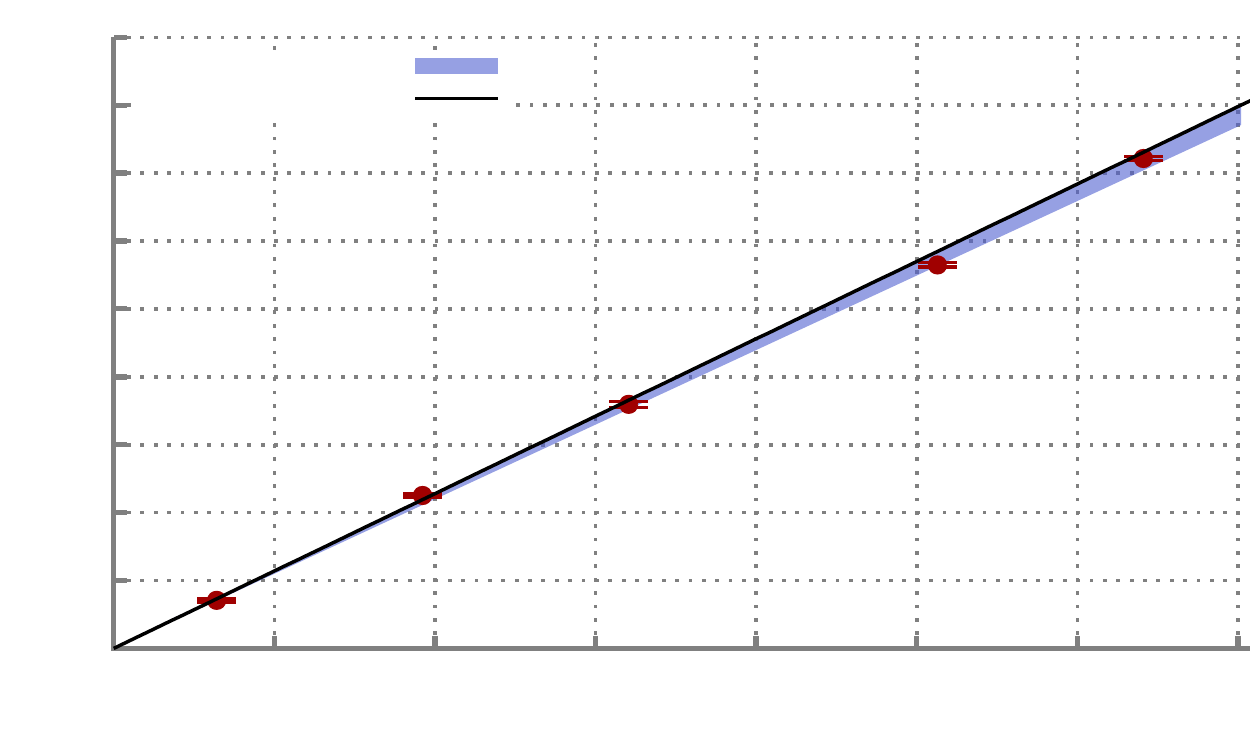}}%
    \gplfronttext
  \end{picture}%
\endgroup